\documentclass{appolb}

\usepackage{graphicx} 
\usepackage{amssymb} 
\usepackage{amstext}
\usepackage{amsfonts} 
\usepackage{slashed} 
\usepackage[dvipsnames]{xcolor}

\newcommand{\be}{\begin{equation}} 
\newcommand{\ee}{\end{equation}}

\begin{document}

\title{Extracting $\sigma_{\pi N}$ from pionic atoms\footnote{Presented 
in June 2019 at the 15th MENU Conf., Pittsburgh~\cite{FG20a}, and at the 
3rd Jagiellonian Symposium on Fundamental and Applied Subatomic Physics, 
Krak\'{o}w~\cite{FG20b}.}} 
\author{Eliahu Friedman and Avraham Gal \address{Racah Institute of Physics, 
The Hebrew University \\ Jerusalem 91904, Israel}} 

\maketitle 

\begin{abstract} 
We discuss a recent extraction of the $\pi N$ $\sigma$ term $\sigma_{\pi N}$ 
from a large-scale fit of pionic-atom strong-interaction data across the 
periodic table. The value thus derived, $\sigma_{\pi N}^{\rm FG}=57\pm 7$~MeV, 
is directly connected via the Gell-Mann--Oakes--Renner expression to the 
medium-renormalized $\pi N$ isovector scattering amplitude near threshold. 
It compares well with the value derived recently by the Bern-Bonn-J\"{u}lich 
group, $\sigma_{\pi N}^{\rm RS}=58\pm 5$~MeV, using the Roy-Steiner equations 
to control the extrapolation of the vanishingly small near threshold $\pi N$ 
isoscalar scattering amplitude to zero pion mass. 
\end{abstract}

\section{Introduction} 
\label{sec:intro} 

\begin{figure}[!t]
\includegraphics[width=0.48\textwidth]{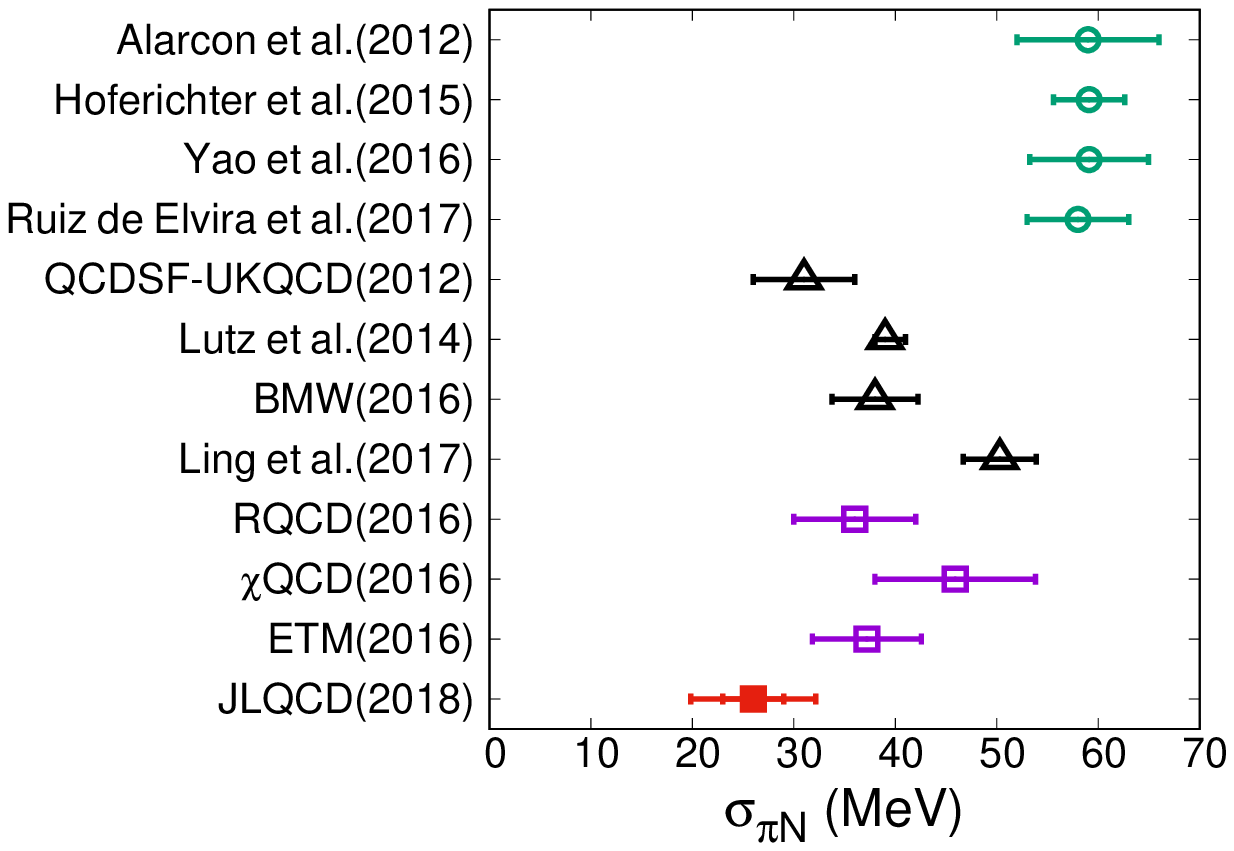} 
\includegraphics[width=0.48\textwidth]{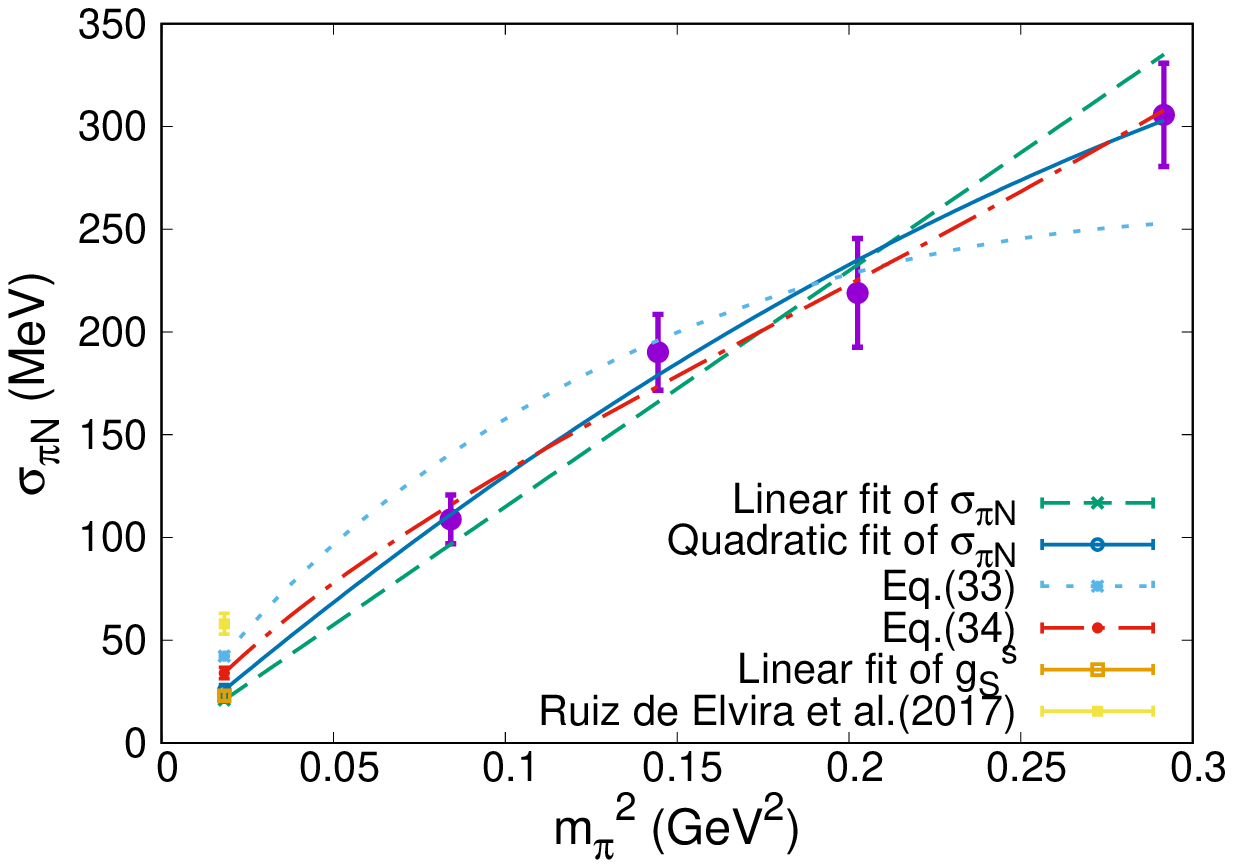} 
\caption{Left: values of $\sigma_{\pi N}$ from recent calculations, 
based on $\pi N$ phenomenology (in green) and in LQCD (other colors). 
Right: chiral extrapolations of LQCD derived $\sigma_{\pi N}$ values to 
the physical pion mass. Figure adapted from Ref.~\cite{JLQCD18}.} 
\label{fig:yam} 
\end{figure}

The $\pi N$ $\sigma$ term 
\begin{equation} 
\sigma_{\pi N}=\frac{{\bar m}_q}{2m_N}\sum_{u,d} \langle N|{\bar q}q|N\rangle, 
\,\,\,\,\,\, {\bar m}_q =\frac{1}{2}(m_u + m_d), 
\label{eq:sigdef} 
\end{equation} 
sometimes called the nucleon $\sigma$ term $\sigma_N$, records the 
contribution of explicit chiral symmetry breaking to the nucleon mass $m_N$ 
arising from the non-zero value of the $u$ and $d$ quark masses in QCD. Early 
calculations yielded a wide range of values, $\sigma_{\pi N}\sim(20-80)$~MeV 
\cite{Sainio02}. Recent calculations use two distinct approaches: 
(i) pion-nucleon low-energy phenomenology guided by chiral EFT, with or 
without solving Roy-Steiner equations, result in values of $\sigma_{\pi N}
\sim(50-60)$~MeV \cite{AMO12,CYZ13,Hof15,DCH16,RdE18}, the most recent of 
which is 58$\pm$5~MeV; and (ii) lattice QCD (LQCD) calculations reach values 
of $\sigma_{\pi N}\sim (30-50)$~MeV \cite{Hor12,BMW16,chiQCD16,ETM16,RQCD16,
JLQCD18,ETM19}, the most recent of which is 41.6$\pm$3.8~MeV. This dichotomy 
is demonstrated on the left panel of Fig.~\ref{fig:yam}. However, when 
augmented by chiral perturbation expansions, LQCD calculations reach also 
values of $\sim$50~MeV, see e.g. Refs.~\cite{LTW00,ALCV13,RGM15,RLG18}. 
Ambiguities in chiral extrapolations of LQCD calculations to the physical 
pion mass are demonstrated on the right panel of Fig.~\ref{fig:yam}. 

A third approach for evaluating $\sigma_{\pi N}$ was recently proposed 
by us~\cite{fg19} focusing on the $\sigma_{\pi N}$-dependent in-medium 
renormalization of the $\pi N$ isovector scattering length $b_1$, 
determined from a wealth of strong-interaction level shifts and widths 
data in pionic atoms across the periodic table~\cite{fg07}. This contrasts 
with extrapolating the vanishigly small $\pi N$ isoscalar scattering length 
$b_0$ from $m_{\pi}\approx 138$~MeV to the Cheng-Dashen point or nearby at 
$m_{\pi}\sim 0$, as done in the first approach. To demonstrate the issues 
involved in comparing these two methodologies, we cite from a recent work 
by the Bern-Bonn-J\"{u}lich group~\cite{Hof16} an expression relating the 
expected departure of the evaluated $\sigma_{\pi N}$ from their value of 
$59\pm 3$~MeV~\cite{Hof15} upon varying the input values of $b_0$ and $b_1$:
\begin{equation} 
\sigma_{\pi N}\approx(59\pm 3)~{\rm MeV} + 1.116\,\Delta b_0^{\rm free} 
+ 0.390\, \Delta b_1^{\rm free}, 
\label{eq:hof16a} 
\end{equation}
where $\Delta b_j^{\rm free}$, $j=0,1$, is the difference between the values 
of $b_j^{\rm free}$ (in units of $10^{-3}\,m_{\pi}^{-1}$) used in a given 
specific model and those used in the calculation of Ref.~\cite{Hof15}. 
Eq.~(\ref{eq:hof16a}) suggests that the uncertainty in the determination of 
$\sigma_{\pi N}$ incurred by the model dependence of $b_0^{\rm free}$ is 
roughly three times larger than that incurred by the model dependence of 
$b_1^{\rm free}$. Regarding the model dependence of these free-space 
scattering lengths we note the two sets of input scattering lengths 
$(b_0^{\rm free},b_1^{\rm free})$ discussed in Ref.~\cite{Hof16}, 
\begin{equation} 
(-0.9,\,-85.3)\times 10^{-3}\,m_{\pi}^{-1},\,\,\,\,\,\,(+7.9,\,-85.4)\times 
10^{-3}\,m_{\pi}^{-1}, 
\label{eq:hof16b} 
\end{equation} 
differing from each other by whether or not charge dependent effects are 
incorporated into the values of scattering lengths derived from $\pi^-$H 
and $\pi^-$d atoms by Baru {\it et al.}~\cite{baru11}. It is evident that 
the charge dependence of the near threshold $\pi N$ interation affects 
dominantly the isoscalar $b_0^{\rm free}$ while leaving the isovector 
$b_1^{\rm free}$ basically intact. This makes an approach based 
on $b_1^{\rm free}$ quite attractive. 

To set the stage for how the third approach works we note that the 
$\pi N$ scattering lengths~\cite{baru11} are well approximated by the 
Tomozawa-Weinberg leading-order (LO) chiral limit~\cite{TWe66} 
\begin{equation} 
b_0^{\rm LO}=0,\;\;\;\;\;\;\; b_1^{\rm LO}=-\frac{\mu_{\pi N}}
{8\pi f^{2}_{\pi}} = -79\times 10^{-3}\,m_{\pi}^{-1}, 
\label{eq:TW} 
\end{equation} 
where $\mu_{\pi N}$ is the $\pi N$ reduced mass and $f_{\pi}=92.2$~MeV 
is the free-space pion decay constant. This expression for the isovector 
amplitude $b_1$ suggests that its in-medium renormalization is directly 
connected to that of $f_{\pi}$, given to first order in the nuclear density 
$\rho$ by the Gell-Mann - Oakes - Renner (GMOR) expression \cite{GMOR68} 
\begin{equation} 
\frac{f_\pi^2(\rho)}{f_\pi^2} = \frac{<\bar q q>_{\rho}}{<\bar q q>} 
\simeq 1 - \frac{\sigma_{\pi N}}{m_{\pi}^2 f_{\pi}^2}\,\rho, 
\label{eq:fpi} 
\end{equation} 
where $<\bar q q>_{\rho}$ stands for the in-medium quark condensate. 
The decrease of $<\bar q q>_{\rho}$ with density in Eq.~(\ref{eq:fpi}) 
marks the leading low-density behavior of the order parameter of the 
spontaneously broken chiral symmetry, see e.g. Ref.~\cite{CFG92}. 
Recalling the $f_{\pi}$ dependence of $b_1^{\rm LO}$ in Eq.~(\ref{eq:TW}), 
Eq.~(\ref{eq:fpi}) suggests the following density dependence for the 
in-medium $b_1$: 
\begin{equation} 
b_1=b_1^{\rm free}\left(1-\frac{\sigma_{\pi N}}{m_{\pi}^2f_{\pi}^2}
\rho\right)^{-1}. 
\label{eq:ddb1} 
\end{equation} 
In this model, introduced by Weise \cite{Wei00,Wei01}, the explicitly 
density-dependent $b_{1}(\rho)$ of Eq.~(\ref{eq:ddb1}) figures directly 
in the pion-nucleus $s$-wave near-threshold potential. Studies of pionic 
atoms~\cite{fg14} and low-energy pion-nucleus scattering~\cite{Fri04,Fri05} 
confirmed that the $\pi N$ isovector $s$-wave interaction term is indeed 
renormalized in agreement with Eq.~(\ref{eq:ddb1}). It is this in-medium 
renormalization that brings in $\sigma_{\pi N}$ to the interpretation of 
pionic-atom data. However, the value of $\sigma_{\pi N}$ was held fixed 
around 50 MeV in these studies, with no attempt to determine its optimal 
value. 

In our recent work~\cite{fg19} we kept to the $\pi N$ isovector $s$-wave 
amplitude $b_1$ renormalization given by Eq.~(\ref{eq:ddb1}), but varied also 
$\sigma_{\pi N}$ in fits to a comprehensive set of pionic atoms data across 
the periodic table. Other real $\pi N$ interaction parameters varied together 
with $\sigma_{\pi N}$ converged at expected free-space values. Holding these 
parameters fixed at the converged values, except for the tiny isoscalar 
$s$-wave single-nucleon amplitude $b_0$ which is renormalized primarily 
by a double-scattering term (see below), we obtained a best-fit value of 
$\sigma_{\pi N}^{\rm FG}=57\pm 7$~MeV. A more comprehensive discussion of our 
fits to pionic atoms data is provided here. The pionic atoms approach used by 
us to extract $\sigma_{\pi N}$ is reviewed in the next section, followed by 
results and discussion in subsequent sections.

\section{Pionic atoms optical potentials}
\label{sec:atom} 

The starting point in our most recent optical-potential analysis of pionic 
atoms~\cite{fg14} is the in-medium pion self-energy $\Pi(E,\vec p,\rho)$ 
that enters the in-medium pion dispersion relation 
\begin{equation} 
E^2-{\vec p}^{~2}-m_{\pi}^2-\Pi(E,\vec p,\rho)=0, 
\label{eq:disp} 
\end{equation} 
where ${\vec p}$ and $E$ are the pion momentum and energy, respectively, 
in nuclear matter of density $\rho$. The resulting pion-nuclear optical 
potential $V_{\rm opt}$, defined by $\Pi(E,\vec p,\rho)=2EV_{\rm opt}$, 
enters the near-threshold pion wave equation 
\begin{equation} 
\left[ \nabla^2  - 2{\mu}(B+V_{\rm opt} + V_c) + (V_c+B)^2\right] \psi = 0, 
\label{eq:KG} 
\end{equation} 
where $\hbar = c = 1$. Here $\mu$ is the pion-nucleus reduced mass, $B$ is 
the complex binding energy, $V_c$ is the finite-size Coulomb interaction of 
the pion with the nucleus, including vacuum-polarization terms, all added 
according to the minimal substitution principle $E \to E - V_c$. Interaction 
terms negligible with respect to $2{\mu}V_{\rm opt}$, i.e. $2V_cV_{\rm opt}$ 
and $2BV_{\rm opt}$, are omitted. We use the Ericson-Ericson form~\cite{EEr66} 
\begin{equation} 
2\mu V_{\rm opt}(r) = q(r) + \vec \nabla \cdot 
\left(\frac{\alpha_1(r)}{1+\frac{1}{3}\xi\alpha_1(r)}+\alpha_2(r)\right)
\vec \nabla, 
\label{eq:EE} 
\end{equation} 
with $s$-wave part $q(r)$ and $p$-wave part, $\alpha_1(r)$ and 
$\alpha_2(r)$, given by \cite{fg07} 
\begin{eqnarray} 
q(r) & = & -4\pi(1+\frac{\mu}{m_N})\{ b_0[\rho_n(r)+\rho_p(r)] 
 +b_1[\rho_n(r)-\rho_p(r)] \} \nonumber \\ 
 & &  -4\pi(1+\frac{\mu}{2m_N})4B_0\rho_n(r) \rho_p(r), 
\label{eq:EEs} 
\end{eqnarray} 
\begin{equation} 
\alpha_1(r) = 4\pi(1+\frac{\mu}{m_N})^{-1}
\{c_0[\tilde{\rho}_n(r)+\tilde{\rho}_p(r)]
+c_1[\tilde{\rho}_n(r)-\tilde{\rho}_p(r)]\}, 
\label{eq:EEp1} 
\end{equation} 
\begin{equation} 
\alpha_2(r) = 4\pi(1+\frac{\mu}{2m_N})^{-1}
4C_0\tilde{\rho}_n(r)\tilde{\rho}_p(r), 
\label{eq:EEp2} 
\end{equation} 
augmented by $p$-wave angle-transformation terms of order $O(m_{\pi}/m_N)$. 
Here $\rho_n$ and $\rho_p$ are neutron and proton density distributions 
normalized to the number of neutrons $N$ and number of protons $Z$, 
respectively, and $\tilde{\rho}_n$ and $\tilde{\rho}_p$ are obtained from 
$\rho_n$ and $\rho_p$ by folding a $\pi N\Delta$ form factor~\cite{gg11}. 
The coefficients $b_0$, $b_1$ in Eq.~(\ref{eq:EEs}) are effective 
density-dependent pion-nucleon isoscalar and isovector $s$-wave scattering 
amplitudes, respectively, evolving from the free-space scattering lengths, 
and are essentially real near threshold. Similarly, the coefficients $c_0$, 
$c_1$ in Eq.~(\ref{eq:EEp1}) are effective $p$-wave scattering amplitudes 
which, since the $p$-wave part of $V_{\rm opt}$ acts mostly near the nuclear 
surface, are close to the free-space scattering volumes provided $\xi =1$ 
is applied in the Lorentz-Lorenz renormalization of $\alpha_1$ in 
Eq.~(\ref{eq:EE}). The parameters $B_0$ and $C_0$ represent multi-nucleon 
absorption and therefore have an imaginary part. Their real parts stand 
for dispersive contributions which often are absorbed into the respective 
single-nucleon amplitudes~\cite{SMa83}. Below we focus on the $s$-wave part 
$q(r)$ of $V_{\rm opt}$. 

Regarding the isoscalar amplitude $b_0$, since the free-space value 
$b_0^{\rm free}$ is exceptionally small, it is customary in the analysis of 
pionic atoms to supplement it by double-scattering contributions induced by 
Pauli correlations. For completeness we also include similar contributions to 
$b_1$ which decrease its value, although by only less than 10\%. Thus, the 
single-nucleon $b_0$ and $b_1$ terms in Eq.~(\ref{eq:EEs}) are extended to 
account also for double-scattering~\cite{EEr66,KEr69}, 
\begin{equation} 
\tilde{b}_0 \rightarrow \tilde{b}_0-\frac{3}{2\pi}\,
(\tilde{b}_0^2+2\tilde{b}_1^2)\,p_F,
\,\,\,\,\,\,\tilde{b}_1 \rightarrow \tilde{b}_1+\frac{3}{2\pi}\,
(\tilde{b}_1^2-2\tilde{b}_0\tilde{b}_1)\,p_F, 
\label{eq:b0b} 
\end{equation} 
where $\tilde{b}_j\equiv (1+\frac{m_\pi}{m_N})b_j$, and $p_F$ is the local 
Fermi momentum corresponding to the local nuclear density $\rho=2p_F^3/
(3\pi^2)$. 

Regarding the isovector amplitude $b_1$, it affects primarily level shifts 
in pionic atoms with $N-Z\neq 0$. However, it affects also $N=Z$ pionic 
atoms through the dominant quadratic $b_1$ contribution to $b_0$ of 
Eq.~(\ref{eq:b0b}). This dominance follows already at the level of $b_1^{\rm 
free}$ from a systematic expansion of the pion self-energy up to $O(p^4)$ 
in nucleon and pion momenta within chiral perturbation theory~\cite{KW01}. 
Following Ref.~\cite{KKW03} it can be argued that it is the in-medium $b_1$ 
Eq.~(\ref{eq:ddb1}) that enters the Pauli-correlation double-scattering 
contribution in Eq.~(\ref{eq:b0b}). This approach has been practised in 
numerous global fits to pionic atoms by us~\cite{fg07,fg14} as well as by 
other groups, e.g. Geissel et al.~\cite{GGG02}, using a fixed value of 
$\sigma_{\pi N}$. To study the role of a variable $\sigma_{\pi N}$ as per 
Eq.~(\ref{eq:ddb1}) we extended $b_1$ wherever appearing in Eq.~(\ref{eq:b0b}) 
by substituting 
\begin{equation} 
b_1 \rightarrow b_1\,\left(1-\frac{\sigma_{\pi N}}{m_{\pi}^2f_{\pi}^2}
\rho\right)^{-1}. 
\label{eq:b1b} 
\end{equation} 

Regarding the nuclear densities $\rho_p$ and $\rho_n$ that enter the 
potential, Eqs.~(\ref{eq:EEs})--(\ref{eq:EEp2}), two-parameter Fermi 
distributions with the same diffuseness parameter for protons and neutrons 
were used~\cite{fg07,Fri09} yielding lower values of $\chi ^2$ than other 
shapes do for pions. With proton densities determined from nuclear charge 
densities, the neutron densities were varied, searching for best agreement 
with the pionic atoms data by assuming a linear dependence of $r_n-r_p$, 
the difference between the root-mean-square (rms) radii, on the neutron excess 
ratio $(N-Z)/A$: 
\begin{equation} 
r_n-r_p = \gamma\, \frac{N-Z}{A} + \delta \; , 
\label{eq:rnrp} 
\end{equation} 
with $\gamma$ close to 1.0~fm and $\delta$ close to zero. Here we used $\delta 
= -0.035$~fm and varied $\gamma$. For example, $\gamma$=1~fm means $r_n-r_p = 
0.177$~fm in $^{208}$Pb, a value compatible with several analyses of pion 
strong and electromagnetic interactions in $^{208}$Pb~\cite{Fri12,MAMI14}, 
and with other determinations of the so called `neutron skin'.

\section{Results} 
\label{sec:res} 

\begin{figure}[htb] 
\begin{center} 
\includegraphics[width=0.48\textwidth]{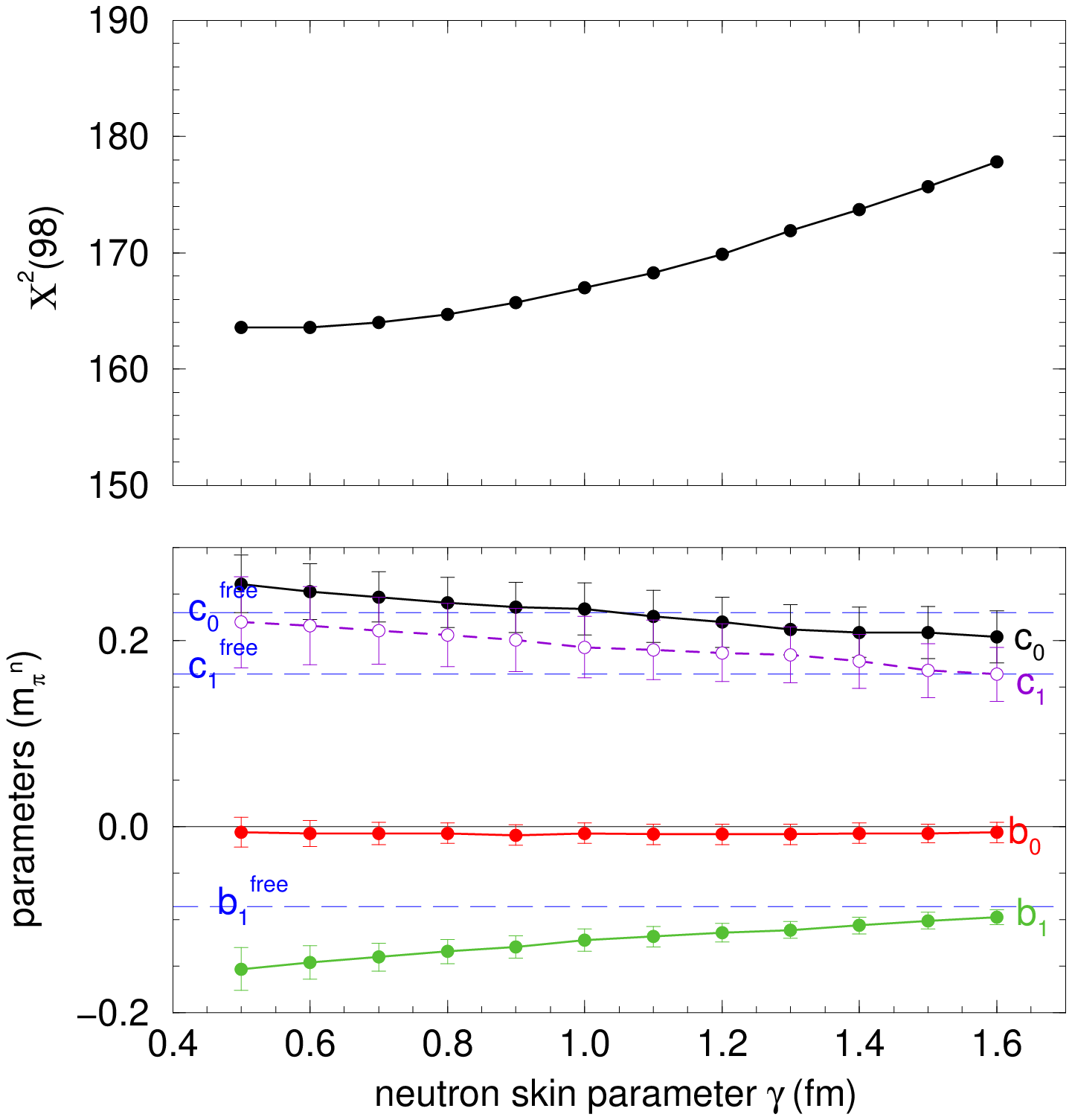} 
\includegraphics[width=0.48\textwidth]{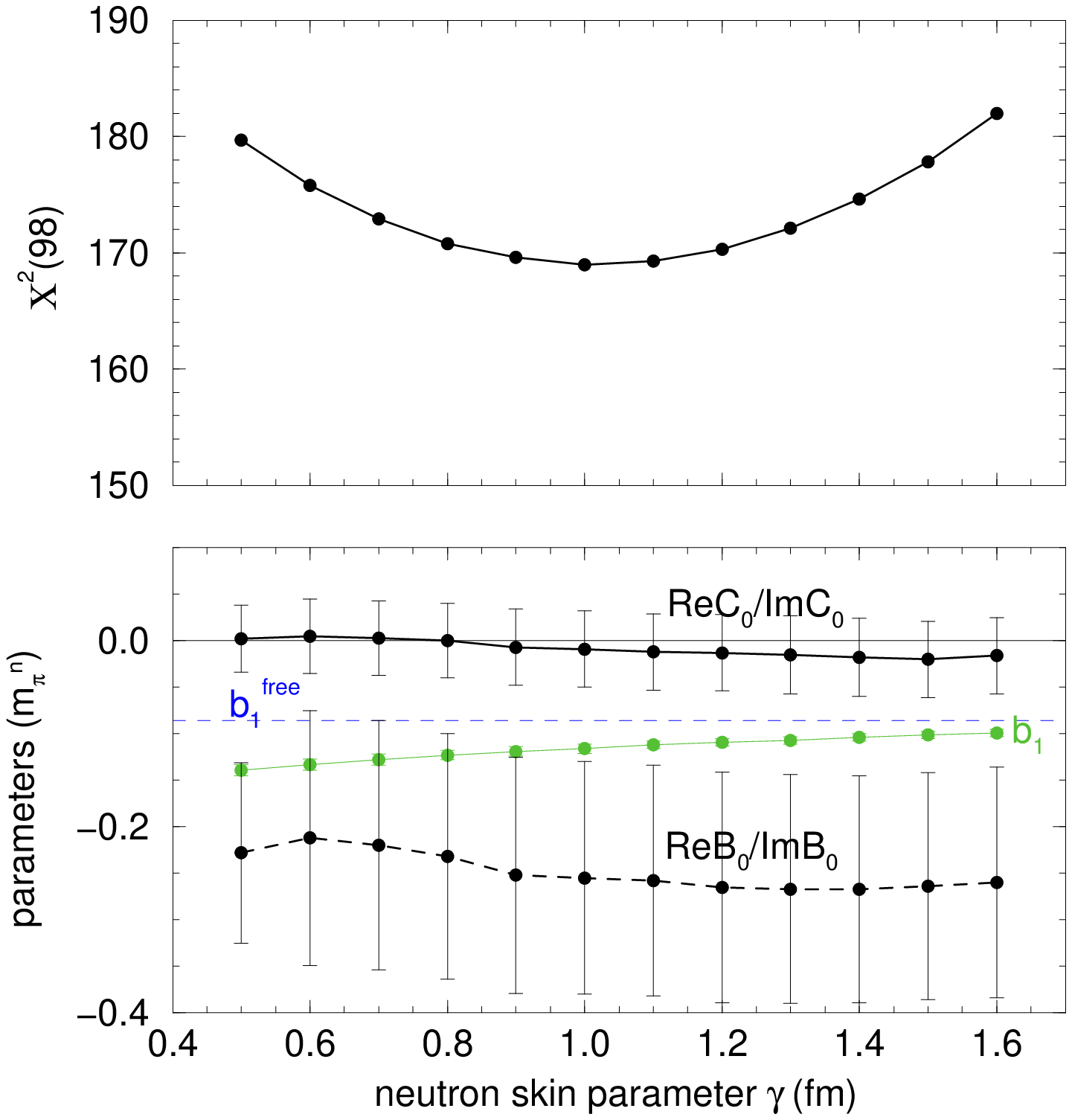} 
\caption{Fits to 98 pionic atoms data points for $\sigma_{\pi N}=0$ as 
a function of the neutron-skin parameter $\gamma$, with $\chi^2$ values 
plotted in the upper panels and fitted values of some of the $\pi$-nucleus 
optical potential parameters plotted in the lower panels. No $\chi^2$ minimum 
is reached in the 8-parameter left-panel fits, but fixing the $p$-wave 
parameters $c_0$ and $c_1$ at their SAID~\cite{SAID06} threshold values 
0.23 and 0.16 $m_{\pi}^{-3}$, respectively, produces the fits shown in 
the right panels.} 
\label{fig:nosigma} 
\end{center} 
\end{figure}

Following the optical potential approach described in the preceding section, 
and more extensively in Refs.~\cite{fg07,fg14}, global fits to strong 
interaction level shifts and widths from Ne to U were made over a wide range 
of values for the neutron-skin parameter $\gamma$ as shown in 
Fig.~\ref{fig:nosigma}. 

The fitted 98 data 
points include `deeply bound' states in Sn isotopes and in $^{205}$Pb. Varying 
all eight parameters (real $b_0$, $b_1$, $c_0$, $c_1$; complex $B_0$, $C_0$) 
in Eqs.~(\ref{eq:EEs})--(\ref{eq:EEp2}) produces good $\chi^2$ fits, $\chi^2 
\sim 170$, but short of a well defined $\chi^2(\gamma)$ minimum as clearly 
seen in the upper left panel of Fig.~\ref{fig:nosigma}. The lower left panel 
shows that the single-nucleon parameters are well determined and vary smoothly 
with $\gamma$. 

Holding the $p$-wave single-nucleon parameters $c_0,c_1$ fixed at their SAID 
free-space threshold values marked by dashed horizontal lines, thereby 
reducing the number of fitted parameters to six, a $\chi^2$ minimum around 
$\gamma = 1$ to 1.1 fm was reached as shown in the upper right panel of 
Fig.~\ref{fig:nosigma}. In these six-parameter fits, Im$\,B_0$ and Im$\,C_0$ 
(not shown) come out well-determined, with values almost independent 
of $\gamma$, but Re$\,B_0$ and Re$\,C_0$ are poorly determined as seen 
in the lower right panel of the figure. In all the fits shown here in 
Fig.~\ref{fig:nosigma}, $b_1$ was treated as a free parameter regardless of 
any possible functional dependence on $\sigma_{\pi N}$, thereby corresponding 
to $\sigma_{\pi N}=0$ in Eq.~(\ref{eq:b1b}). The fitted values of $b_1$ 
disagree then over a broad range of $\gamma$s with the value $b_1^{\rm free}$ 
marked by a dashed horizontal line. 

\begin{figure}[htb] 
\begin{center} 
\includegraphics[width=0.48\textwidth]{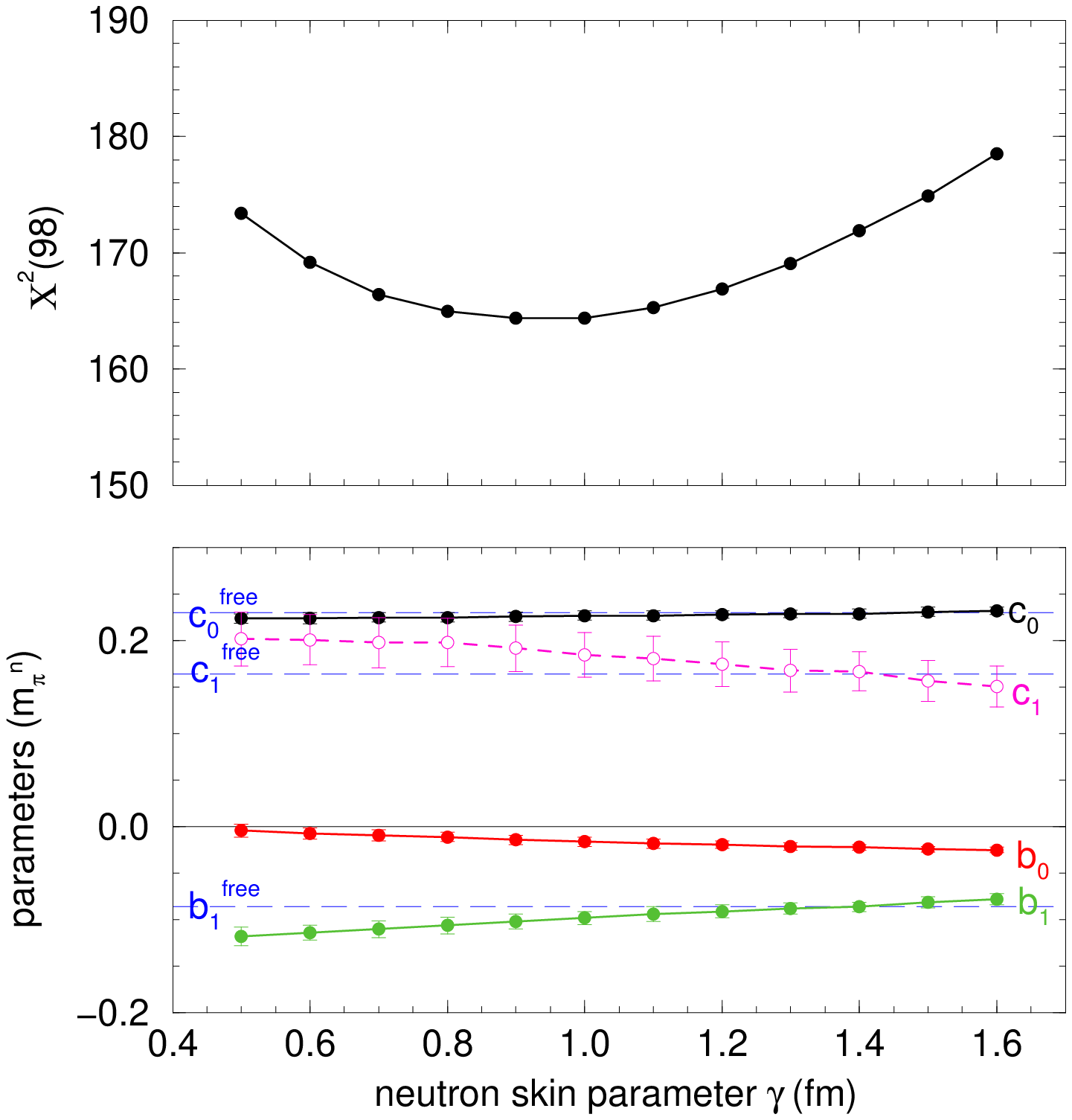} 
\includegraphics[width=0.48\textwidth]{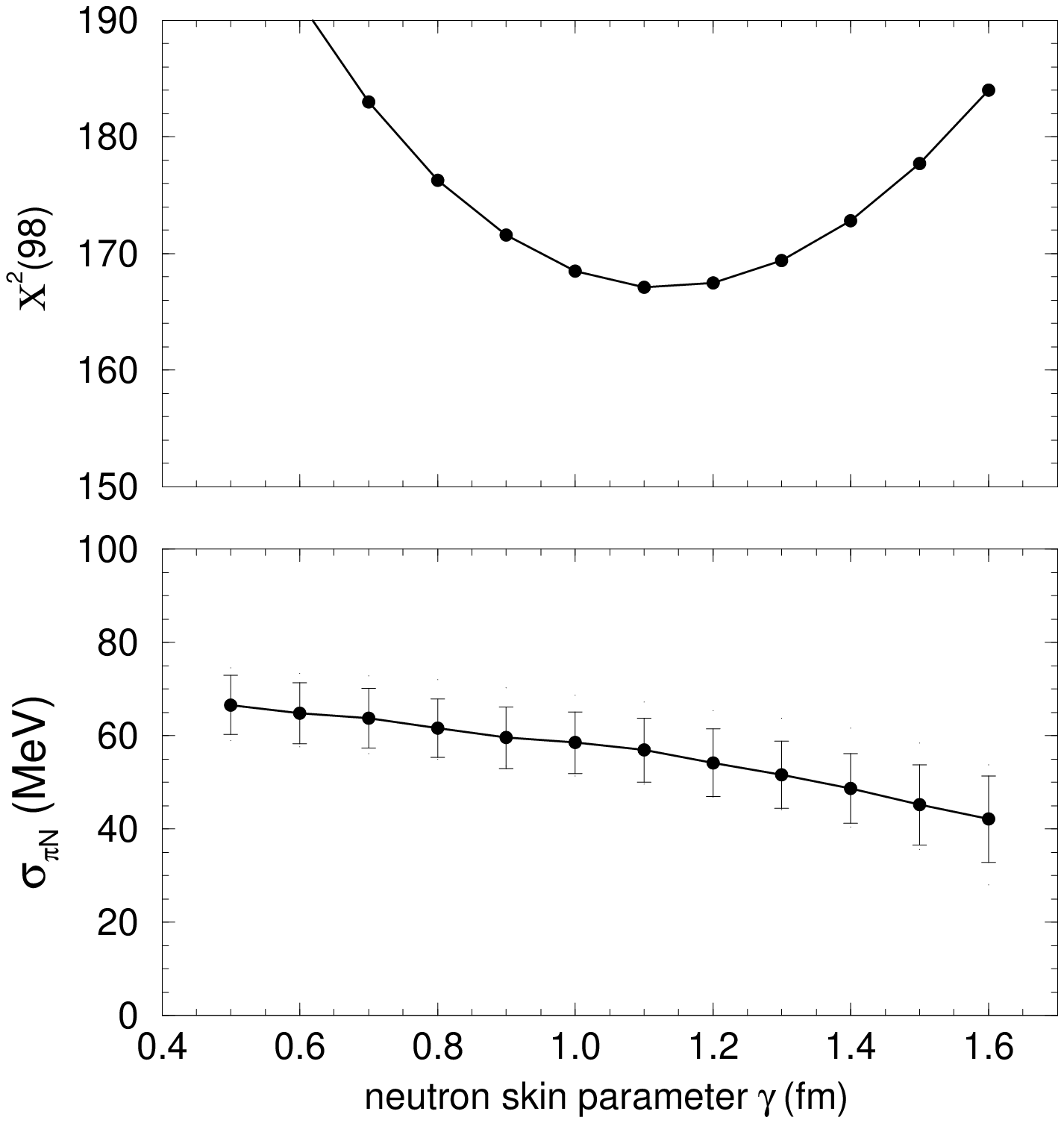} 
\caption{Left: 6-parameter fits with $\sigma_{\pi N}=50$~MeV where Re$\,B_0$ 
and Re$\,C_0$ are kept zero. Right: 4-parameter fits where $c_0$ and 
$c_1$, additionally, are kept at their SAID~\cite{SAID06} threshold values 
0.23 and 0.16~$m_{\pi}^{-3}$, respectively. Of the 4 varied parameters 
($b_0$, $b_1$, Im$\,B_0$, Im$\,C_0$) $b_1$ is related to $\sigma_{\pi N}$ by 
Eq.~(\ref{eq:ddb1}). Resulting values of $\sigma_{\pi N}$ are plotted in the 
lower right panel.}
\label{fig:50sigma}
\end{center} 
\end{figure}

Introducing the in-medium density dependence of $b_1$ given by 
Eq.~(\ref{eq:b1b}) in terms of $\sigma_{\pi N}$ we first demonstrate the 
effect of using a fixed value of $\sigma_{\pi N}=50$~MeV, as practised in 
all of our past works~\cite{fg14}, on the fitted parameters. This is shown 
within six-parameter fits in the left panels of Fig.~\ref{fig:50sigma}. 
Rather than keeping the $p$-wave single-nucleon parameters $c_0$ and $c_1$ to 
their SAID free-space threshold values, as done in the $\sigma_{\pi N}=0$ fits 
shown in the right panels of Fig.~\ref{fig:nosigma}, here we kept Re$\,B_0$ 
and Re$\,C_0$ to zero values thereby producing as good fits to the data as by 
letting them vary. In particular, suppressing Re$\,B_0$ in pionic atoms fits 
amounts to absorbing it into an effective $b_0$ parameter~\cite{SMa83}. The 
fitted $c_0$ and $c_1$, particularly $c_0$, are clearly seen in the lower 
panel to come out close to the respective free-space values. As for the 
$s$-wave single-nucleon parameters $b_0$ and $b_1$, the dominance of $b_1$ 
with respect to $b_0$ is also clearly seen. The introduction of a nonzero 
value of $\sigma_{\pi N}$ allows $b_1$ to reach its free-space value 
$b_1^{\rm free}$ beginning at a neutron-skin parameter $\gamma$ value of 
1.1~fm. 

Holding now the $p$-wave single-nucleon parameters $c_0$ and $c_1$ at their 
free-space SAID threshold values 0.23 and 0.16~$m_{\pi}^{-3}$, respectively, 
and keeping as before Re$\,B_0$ and Re$\,C_0$ to zero values, we show in the 
right panels of Fig.~\ref{fig:50sigma} four-parameter fits where the varied 
parameters are $b_0$, $\sigma_{\pi N}$ for $b_1$ using Eq.~(\ref{eq:ddb1}), 
Im$\,B_0$ and Im$\,C_0$. A minimum value of $\chi^2_{\rm min}=167.1$ 
is reached at $\gamma\approx 1.1$~fm where $\sigma_{\pi N}$ assumes 
a value of $\sigma_{\pi N}^{\rm FG}=56.9\pm 6.9$~MeV. Note that a value of 
$\gamma\approx 1.1$~fm agrees with other determinations of this quantity in 
$^{208}$Pb~\cite{MAMI14}. To check the dependence of $\sigma_{\pi N}$ on $b_0$ 
we repeated fits with $b_0$ kept fixed at either one of the two free-space 
threshold values listed in Eq.~(\ref{eq:hof16b}), varying then also Re$\,B_0$ 
and Re$\,C_0$. Typical $\chi^2$ values increased by 20 to 30, but the $\chi^2$ 
minima remained at $\gamma=1.1$ to 1.2~fm with corresponding values of 
$\sigma_{\pi N}$ decreasing at most by 3~MeV. We also note that the resulting 
value of $\sigma_{\pi N}$ is identical with that derived in our recently 
published work~\cite{fg19} where the effect on the derived value of 
$\sigma_{\pi N}$ of form-factor folding, $\rho_{n,p}\rightarrow 
\tilde{\rho}_{n,p}$ in the $p$-wave terms (\ref{eq:EEp1},\ref{eq:EEp2}) 
of the pion-nucleus optical potential, was shown to be negligibly small.

\section{Discussion and summary} 
\label{sec:disc} 

The pionic atoms fits and the value of the $\pi N$ $\sigma$ term
$\sigma_{\pi N}$ extracted in the present work are based on the in-medium
renormalization of the near-threshold $\pi N$ isovector scattering amplitude
$b_1$ as given by Eq.~(\ref{eq:ddb1}), derived at LO from Eqs.~(\ref{eq:TW}) 
and (\ref{eq:fpi}) for the in-medium decrease of the pion decay constant 
$f_{\pi}$ associated via the GMOR expression with the in-medium decrease of 
the quark condensate $<\bar q q>$. Higher order corrections to this simple 
form have been proposed in the literature and were discussed by us in 
Ref.~\cite{fg19}. Briefly, one may classify two such corrections arising from: 
(i) $NN$ correlation contributions~\cite{KHW08} from one- and two-pion 
interaction terms, increasing the fitted $\sigma_{\pi N}$ value by about 7 MeV 
(or by a smaller amount following a chiral approach at NLO~\cite{LOM10}); and 
(ii) an upward shift of the in-medium pion mass $m_{\pi}(\rho)$ in symmetric 
nuclear matter from its free-space value~\cite{JHK08}, decreasing the fitted 
$\sigma_{\pi N}$ value by a similar amount, and also by adding corrections 
of order $\rho^{4/3}$~\cite{GJi13,GJi14} which at a typical nuclear density 
$\rho_{\rm eff}=0.1$~fm$^{-3}$~\cite{SMa83} are negligible. Interestingly 
but perhaps fortuitously, these two higher-order effects largely cancel 
each other. 

In conclusion, we have derived in this work a value of $\sigma_{\pi N}^{\rm FG}
=57\pm 7$~MeV from a large scale fit to pionic atoms observables, in agreement 
with the relatively high $\sigma_{\pi N}$ values reported in recent studies 
based on modern hadronic $\pi N$ phenomenology~\cite{RdE18}, but in 
disagreement with the considerably lower $\sigma_{\pi N}$ values reached 
in some of the recent modern lattice QCD calculations, e.g.~\cite{JLQCD18}. 
Our derivation is based on the model introduced by Weise and 
collaborators~\cite{Wei00,Wei01,KKW03} for the in-medium renormalization 
of the $\pi N$ near-threshold isovector scattering amplitude, using its 
leading density dependence Eq.~(\ref{eq:ddb1}), and was found robust in 
fitting the wealth of pionic atoms data against variation of other $\pi N$ 
interaction parameters that enter the low-energy pion self-energy operator. 
The two types of model corrections beyond the leading density dependence 
considered here were found to be relatively small, a few MeV each, and partly 
canceling each other. Further model studies are desirable in order to confirm 
this conclusion.

\section*{Acknowledgments} 
We are grateful to Norbert Kaiser, Wolfram Weise and Nodoka Yamanaka 
for useful correspondence on the subject of the present work.


\begin{thebibliography}{99} 

\bibitem{FG20a} E.~Friedman, A.~Gal, AIP Conf. Proc. \textbf{2249}, 030015 
(2020). 

\bibitem{FG20b} E.~Friedman, A.~Gal, Acta Phys. Pol. B \textbf{51}, 45 (2020). 

\bibitem{Sainio02} M.E.~Sainio, $\pi N$ Newsletter \textbf{16}, 138 (2002) 
[arXiv:hep-ph/0110413]. 

\bibitem{AMO12} J.M.~Alarc\'{o}n, J.M.~Camalich, J.A.~Oller, 
Phys. Rev. D \textbf{85}, 051503 (2012). 

\bibitem{CYZ13} Y.-H.~Chen, D.-L.~Yao, H.Q.~Zheng, Phys. Rev. D \textbf{87}, 
054019 (2013). 

\bibitem{Hof15} M.~Hoferichter, J.~Ruiz de Elvira, B.~Kubis, 
U.-G. Mei{\ss}ner, Phys. Rev. Lett. \textbf{115}, 092301 (2015). 

\bibitem{DCH16} V. Dmitra\v{s}inovi\'{c}, H.-X.~Chen, A.~Hosaka, 
Phys. Rev. C \textbf{93}, 065208 (2016). 

\bibitem{RdE18} J.~Ruiz de Elvira, M.~Hoferichter, B.~Kubis, 
U.-G. Mei{\ss}ner, J. Phys. G \textbf{45}, 024001 (2018).

\bibitem{Hor12} R.~Horsley {\it et al.} (QCDSF-UKQCD Collab.), 
Phys. Rev. D \textbf{85}, 034506 (2012). 

\bibitem{BMW16} S.~Durr {\it et al.} (BMW Collab.), Phys. Rev. Lett. 
\textbf{116}, 172001 (2016). 

\bibitem{chiQCD16} Y.-B.~Yang, A.~Alexandru, T.~Draper, K.-F.~Liu 
($\chi$QCD Collab.), Phys. Rev. D \textbf{94}, 054503 (2016). 

\bibitem{ETM16} A.~Abdel-Rehim {\it et al.} (ETM Collab.), 
Phys. Rev. Lett. \textbf{116}, 252001 (2016). 

\bibitem{RQCD16} G.S.~Bali {\it et al.} (RQCD Collab.), 
Phys. Rev. D \textbf{93}, 094504 (2016). 

\bibitem{JLQCD18} N.~Yamanaka, S.~Hashimoto, T.~Kaneko, H.~Ohki 
(JLQCD Collab.), Phys. Rev. D \textbf{98}, 054516 (2018). 

\bibitem{ETM19} C.~Alexandrou {\it et al.} (ETM Collab.), 
arXiv:1909.00485v1. 

\bibitem{LTW00} D.B.~Leinweber, A.W.~Thomas, S.V.~Wright, Phys. Lett. 
B \textbf{482}, 109 (2000). 

\bibitem{ALCV13} L.~Alvarez-Ruso, T.~Ledwig, J.M.~Camalich, 
M.J.~Vicente-Vacas, Phys. Rev. D \textbf{88}, 054507 (2013).

\bibitem{RGM15} X.-L.~Ren, L.-S.~Geng, J.~Meng, Phys. Rev. D 
\textbf{91}, 051502(R) (2015).

\bibitem{RLG18} X.-L.~Ren, X.-Z.~Ling, L.-S.~Geng, Phys. Lett. B 
\textbf{783}, 7 (2018). 

\bibitem{fg19} E.~Friedman, A.~Gal, Phys. Lett. \textbf{792}, 340 (2019). 

\bibitem{fg07} E.~Friedman, A.~Gal, Phys. Rep. \textbf{452}, 89 (2007). 

\bibitem{Hof16} M.~Hoferichter, J.~Ruiz de Elvira, B.~Kubis, 
U.-G. Mei{\ss}ner, Phys. Lett. B \textbf{760}, 74 (2016).

\bibitem{baru11} V.~Baru {\it et al.}, Phys. Lett. B \textbf{694}, 473 (2011); 
Nucl. Phys. A \textbf{872}, 69 (2011). 

\bibitem{TWe66} Y.~Tomozawa, Nuovo Cim. A \textbf{46}, 707 (1966); 
S.~Weinberg, Phys. Rev. Lett. \textbf{17}, 616 (1966).

\bibitem{GMOR68} M.~Gell-Mann, R.J.~Oakes, B.~Renner, Phys. Rev. \textbf{175}, 
2195 (1968).

\bibitem{CFG92} T.D.~Cohen, R.J.~Furnstahl, D.K.~Griegel, Phys. Rev. C 
\textbf{45}, 1881 (1992). 

\bibitem{Wei00}W.~Weise, Acta Phys. Pol. B \textbf{31}, 2715 (2000). 

\bibitem{Wei01}W.~Weise, Nucl. Phys. A \textbf{690}, 98c (2001). 

\bibitem{fg14} E.~Friedman, A.~Gal, Nucl. Phys. A \textbf{928}, 128 (2014), 
and references therein to earlier work on pionic atoms. 

\bibitem{Fri04} E.~Friedman {\it et al.}, Phys. Rev. Lett. \textbf{93}, 
122302 (2004). 

\bibitem{Fri05} E.~Friedman {\it et al.}, Phys. Rev. C \textbf{72}, 034609 
(2005). 

\bibitem{EEr66} M.~Ericson, T.E.O.~Ericson, Ann. Phys. \textbf{36}, 323 
(1966).

\bibitem{gg11} A.~Gal, H.~Garcilazo, Nucl. Phys. A \textbf{864}, 153 
(2011); see Fig. 2 \& Tab. 3. 

\bibitem{SMa83} R.~Seki, K.~Masutani, Phys. Rev. C \textbf{27}, 2799 
(1983). 

\bibitem{KEr69} M.~Krell, T.E.O.~Ericson, Nucl. Phys. B \textbf{11}, 521 
(1969).

\bibitem{KW01} N.~Kaiser, W.~Weise, Phys. Lett. B \textbf{512}, 283 (2001).

\bibitem{KKW03} E.E.~Kolomeitsev, N.~Kaiser, W.~Weise, Phys. Rev. Lett. 
\textbf{90}, 092501 (2003). 

\bibitem{GGG02} H.~Geissel {\it et al.}, Phys. Lett. B \textbf{549}, 
64 (2002); Phys. Rev. Lett. \textbf{88}, 122301 (2002).

\bibitem{Fri09} E.~Friedman, Hyperfine Interact. \textbf{193}, 33 (2009).

\bibitem{Fri12} E.~Friedman, Nucl. Phys. A \textbf{896}, 46 (2012). 

\bibitem{MAMI14} C.M.~Tarbert {\it et al.}, Phys. Rev. Lett. \textbf{112}, 
242502 (2014). 

\bibitem{SAID06} R.A.~Arndt, W.J.~Briscoe, I.I.~Strakovsky, R.L.~Workman, 
Phys. Rev. C \textbf{74}, 045205 (2006); evolving SAID program 
http://gwdac.phys.gwu.edu/ 

\bibitem{KHW08} N.~Kaiser, P.~de Homont, W.~Weise, Phys. Rev. C 
\textbf{77}, 025204 (2008). 
      
\bibitem{LOM10} A.~Lacour, J.A.~Oller, U.-G. Mei{\ss}ner, J. Phys. G 
\textbf{37}, 125002 (2010). 

\bibitem{JHK08} D.~Jido, T.~Hatsuda, T.~Kunihiro, Phys. Lett. B 
\textbf{670}, 109 (2008). 

\bibitem{GJi13} S.~Goda, D.~Jido, Phys. Rev. C \textbf{88}, 065204 (2013). 

\bibitem{GJi14} S.~Goda, D.~Jido, Prog. Theor. Exp. Phys. \textbf{2014}, 
033D03 (2014). 


\end{thebibliography}
\end{document}